\documentstyle[12pt,twoside]{article} 
\normalbaselines 
\font\tbf = cmbx12 
 
\begin{document}

\indent 
\vskip 1cm 
\centerline{\tbf OPTICAL ACTIVITY INDUCED BY CURVATURE} 
\vskip 0.3cm 
\centerline{\tbf IN A GRAVITATIONAL PP-WAVE BACKGROUND} 
\vskip 0.3cm 
\centerline{by}{} 
\vskip 0.3cm 
\centerline{\tbf Alexander B. Balakin } 
\vskip 0.3cm 
\centerline{\it Kazan State University, Kremlevskaya street 18, 
420008, Kazan,  Russia,} 
\centerline{\it e-mail: Alexander.Balakin@ksu.ru} 
\vskip 0.5cm 
\centerline{and} 
\vskip 0.5cm 
\centerline{\tbf Jos\'e P. S. Lemos } 
\vskip 0.3cm 
\centerline{\it Centro Multidisciplinar de Astrof\'{\i}sica -- CENTRA,} 
\centerline{\it Departamento de F\'{\i}sica, Instituto Superior 
T\'ecnico,} 
\centerline{\it Av. Rovisco Pais 1, 1049-001 Lisboa, Portugal,} 
\centerline{\it and} 
\centerline{\it Observat\'orio Nacional - MCT,} 
\centerline{\it Rua General Jos\'e Cristino 77, 20921, 
Rio de Janeiro, Brazil,} 
\centerline{\it e-mail: lemos@kelvin.ist.utl.pt} 
\vskip 3cm 
{\tbf Abstract} 
\indent 
{\small 
We study optical activity induced by curvature. 
The optical activity model we present has 
two phenomenological gyration parameters, within which we analyze three 
model cases, namely, an exactly integrable model, the Landau-Lifshitz 
model and the Fedorov model, these latter two are solved in the short 
wavelength approximation. The model background is a gravitational 
pp-wave.  The solutions show that the optical activity induced by 
curvature leads to Faraday rotation.} 
 
\newpage 
 
\section{Introduction} 
 
The simplest instances of the phenomenon of optical activity, or 
gyrotropy, are (i) the Faraday rotation, i.e., the rotation of the 
polarization plane for a linearly polarized electromagnetic wave 
propagating in a medium, and (ii) birefringence, for left-hand and 
right-hand circularly polarized electromagnetic waves 
\cite{landau1}.  Optical activity appears in materials 
which show linear spatial dispersion of light, as in the case of some 
crystals \cite{landau1}.  Indeed, of the 32 symmetry classes of 
crystals only 15 admit optical activity, and have been studied 
theoretically and experimentally \cite{wood}.  Phenomenological models 
of optical activity \cite{condon}-\cite{mcclain} 
are of great importance in applications to optics 
\cite{marzlin}.

Following Landau and Lifshitz \cite{landau2}, the gravitational field 
in vacuum can be considered as a specific sort of medium.  An example 
of this, and as first shown by Drummond and Hathrell \cite{drummond}, 
vacuum interacting with curvature produces birefringence for an 
electromagnetic wave propagating in a curved spacetime. A glamourous 
consequence, is the gravitational rainbow considered by Lafrance and 
Myers \cite{lafrance}. Other effects related to quantum 
electrodynamics in background fields (either gravitational or 
electromagnetic) show that the propagation of light depends on various 
factors, such as, photon polarization, photon frequency (temporal 
dispersion), direction of propagation, and spacetime location 
\cite{sch}-\cite{dittrich}. 
As has been accepted recently (see, e.g. \cite{dalvit} and references 
therein), the one-loop gravitationally corrected Maxwell equations 
appear to be mathematically equivalent to the classical electrodynamic 
equations for the model of a linear, anisotropic, nonhomogeneous 
media. In this context the influence of the curvature on the 
electromagnetic waves has been described in terms of spacetime 
dependent electric and magnetic permeability tensors 
(see \cite{dalvit} and \cite{bala}-\cite{bale2}). 
 
In this paper, we focus on the phenomenon of optical activity induced 
by the curvature of spacetime. 
First, our description of optical activity is phenomenological.  The 
interaction constants, i.e., the gyration parameters, considered in 
our model are not derived from quantum electrodynamics.  Nevertheless, 
the approach is motivated by the same principles, related to quantum 
corrections to Maxwell equations in a curved background 
\cite{sch}-\cite{dittrich}.  The modeling of optical activity is of 
interest not only from a microscopic point of view, but also as a 
phenomenological set up \cite{nieves}. 
Second, we choose a pp-wave as the background spacetime. It is a 
background well suited for our purposes, since it is an exact 
solution of Einstein field equations and has an explicit time 
dependence. It has attracted much attention in the recent literature 
in connection to the fact that it is the Penrose limit of any spacetime 
\cite{penrose}. 
The main goal of the paper is to present an exact 
solvable model describing the optical activity induced by curvature, 
along the line of papers \cite{bala}-\cite{bale2}. The results show 
that the interaction of the electromagnetic field with the background 
curvature (a pp-wave here) produces a Faraday rotation of the 
polarization vector, as in the classical theory of optical 
activity. The rotation frequency happens to be a function of the 
electromagnetic wave frequency and the Riemann tensor.

\section{Two-parameter model of optical activity} 
 
\subsection{The three-dimensional classical prototype} 
 
Maxwell equations in a dielectric medium  with vanishing current of free 
charges are 
\begin{equation} 
\vec{\nabla} \cdot \vec{D} = 0\,, \quad 
\vec{\nabla} \cdot \vec{B} = 0\,, \quad 
\vec{\nabla} \times \vec{E} = - 
\frac{ \ \partial \vec{B} }{c \partial t}\,, \quad 
\vec{\nabla} \times \vec{H} = 
\frac{ \ \partial \vec{D} }{c \partial t}\,. 
\label{3max} 
\end{equation} 
These equations include as usual the three-vectors $\vec{D}$ and 
$\vec{B}$ (the electric and magnetic induction vectors, respectively), 
as well as the three-vectors $\vec{E}$ and $\vec{H}$ (the electric and 
magnetic fields, respectively).  $\vec{\nabla}$ is the standard 
three-dimensional differential vector operator. In the framework of 
linear electrodynamics, Landau and Lifshitz \cite{landau1} 
connect the $\vec{D}$ and $\vec{E}$ three-vectors as follows, 
\begin{equation} 
D^a = \varepsilon^{ab} E_b + \gamma^{abc} \partial_b E_c \,, 
\label{3constitutive} 
\end{equation} 
where $\varepsilon^{ab}$ is the electric permeability tensor, 
$\gamma^{abc}$ is the optical activity tensor of rank three (it 
describes the gyration properties of the material), 
and $\partial_b$ denotes partial derivative. In this subsection 
latin indices $a,\,b,\,c$ are Euclidean indices running from 1 to 3. 
When the medium is isotropic or has cubic symmetry, the $\gamma^{abc}$ 
tensor reduces to a pseudo-scalar $f$ 
\begin{equation} 
\gamma^{abc} = \frac{c}{\omega} f e^{abc}\,, 
\label{gamma} 
\end{equation} 
where $\omega$ is a constant with units of frequency, and $e^{abc}$ is 
the three-dimensional totally skew Levi-Civita symbol 
\cite{landau1}. The pseudo-scalar $f$ is the gyration 
parameter (it has no units).  Thus, for the case of isotropic medium 
the relation (\ref{3constitutive}) has the three-vector form 
\begin{equation} 
\vec{D} = \varepsilon \vec{E} + 
\frac{cf}{\omega} \vec{\nabla} \times \vec{E} , 
\label{drote} 
\end{equation} 
where the standard definition of the vector product 
$\left[\vec{\nabla} \times \vec{E}\right]^a \equiv e^{abc} \partial_b E_c$ 
has been used, as well as the representation of the  electric permeability 
tensor $\varepsilon_{ab}= \varepsilon \delta_{ab}$, which is valid for the 
isotropic case.  $\delta_{ab}$ is a Kronecker delta with Euclidean indices. 
 
When one puts the standard condition $B^a = \mu^{ab} H_b$, 
where $\mu^{ab}$ is the magnetic permeability, one 
obtains the so-called non-symmetric Landau-Lifshitz's model for optical 
activity (see, e.g. \cite{mcclain}). 
More generally, one can represent the dependence of $\vec{B}$ in $\vec{H}$ 
by the equation, (see, e.g. \cite{fedorov,mcclain}) 
\begin{equation} 
B^a = \mu^{ab} H_b + \bar{\gamma}^{abc} \partial_b H_c \,, 
\label{33constitutive} 
\end{equation} 
where the tensor $\bar{\gamma}^{abc}$ describes the magnetic aspects of the 
optical activity. Analogously to the relation (\ref{drote}), in 
an isotropic medium one obtains 
\begin{equation} 
\vec{B} = \mu \vec{H} + 
\frac{c \bar{f}}{\omega} \vec{\nabla} \times \vec{H} \,, 
\label{hrotb} 
\end{equation} 
where $\mu$ is the magnetic permeability of the isotropic medium and $\bar{f}$ 
is a new parameter, playing the same role for the magnetic field as $f$ plays 
for the electric one. 
Finally, using the simplest model for the 
description of the properties of optically active media 
(see, e.g. \cite{fedorov,marzlin}), one can replace $\varepsilon$ and $\mu$ 
by their values in vacuum  ($\varepsilon=1$ and $\mu=1$). 
Introducing two new gyration parameters $a$ and $b$, defined by 
$a \equiv \frac{c f}{\omega}$ and  $b \equiv -\frac{c\bar{f}}{\omega}$, 
we obtain the three-dimensional form of the two-parameter model describing 
the optical activity, 
\begin{equation} 
\vec{D} =  \vec{E} + a \ \vec{\nabla} \times \vec{E} \,, \quad 
\vec{B} =  \vec{H} - b \ \vec{\nabla} \times \vec{H} \,. 
\label{3model} 
\end{equation} 
It will be better, for what follows, when we discuss the covariant 
generalization of equations (\ref{3model}), to have $\vec{H}$ as a 
function of $\vec{B}$.  Thus, inverting the second equation in 
(\ref{3model}) and keeping only up to first order terms in $b$ one 
obtains 
\begin{equation} 
\vec{D} =  \vec{E} + a \ \vec{\nabla} \times \vec{E} \,, \quad 
\vec{H} =  \vec{B} + b \ \vec{\nabla} \times \vec{B} \,. 
\label{3model2} 
\end{equation} 
This two-parameter set of equations has three important one-parameter 
special cases. For $a=b$ one obtains a model showing 
optical activity, which is fully integrable in the covariant analysis 
in a gravitational pp-wave background, (see section 3.3.1). 
For $b=0$ and $a=-b$ one obtains models not fully integrable but 
which can be dealt in the short wavelength approximation. 
The  $b=0$ model is called the Landau-Lifshitz model (see equation 
(\ref{3constitutive}) and paragraph before equation 
(\ref{33constitutive}), see also the first part of section 3.3.2). 
The $a=-b$ model, or Fedorov model, is 
very special indeed, since the constitutive relations preserve the 
duality invariance between $(\vec{E},\vec{H})$ and $(\vec{D},\vec{B})$ 
already present in the Maxwell equations in the non-conducting and 
non-charged media when one also includes magnetic monopole source 
terms \cite{jackson} (see the second part of section 3.3.2). 
 
\subsection{Covariant representation of the  three-dimensional classical 
prototype in Minkowski spacetime} 
 
If we want to have a covariant model of optical activity we should rely on 
the covariant equations of electrodynamics of continuous media \cite{landau1}, 
\begin{equation} 
\nabla_{k} H^{ik} = 0 \,, 
\label{maxwell1} 
\end{equation} 
\begin{equation} 
\nabla_{k} F^{*ik} =0 \,, 
\label{maxwell2} 
\end{equation} 
where $\nabla_{k}$ is the covariant derivative, 
$H^{ik}$ is the electric-magnetic induction tensor, 
and $F^{*ik}$ is the tensor dual to 
the Maxwell tensor $F_{mn}$, i.e., 
\begin{equation} 
F^{*ik} = \frac{1}{2 \sqrt{-g}}\epsilon^{ikls} F_{ls}\,. 
\label{dual} 
\end{equation} 
Here, $\frac{1}{\sqrt{-g}}\epsilon^{ikls}$ is the four-dimensional 
Levi-Civita tensor and 
$\epsilon^{ikls}$ is the completely anti-symmetric four-dimensional symbol 
with $\epsilon^{0123} = - \epsilon_{0123} = 1$. 
Latin indices $i,\,j,\,k,\,...$ run from 0 to 3. 
 
The constitutive equations, connecting the induction tensor $H^{ik}$ with the 
Max\-well tensor $F_{mn}$ can be written in the following form 
\begin{equation} 
H^{ik} = C^{ikmn} F_{mn} + \nabla_s (D^{iksmn} F_{mn}) \,, 
\label{constitutive} 
\end{equation} 
where the so-called material tensor $C^{ikmn}$ describes the 
properties of the linear response  (it includes the 
electric and magnetic permeabilities, as well as the magneto-electric 
coefficients), and the $D^{iksmn}$ tensor includes the coefficients of 
the linear dispersion.  These tensors have similar symmetry properties 
with respect to indices transposition, namely, 
\begin{equation} 
C^{ikmn}= - C^{kimn}= - C^{iknm} = C^{mnik} \,, 
\label{symmetry1} 
\end{equation} 
\begin{equation} 
D^{iksmn}= - D^{kismn} = - D^{iksnm} \,. 
\label{symmetry} 
\end{equation} 
Nevertheless, the $D^{iksmn}$ tensor is not in general symmetric with 
respect to the transposition of the pairs $(ik)$ and $(mn)$.  The last 
term on the right hand side of equation (\ref{constitutive}) is a 
covariant generalization of the three-dimensional constitutive 
equations (\ref{3constitutive}) and (\ref{33constitutive}). 
 
In order to construct explicitly the tensor $D^{iksmn}$ for the 
two-parameter model of the optical activity we recall, first, that 
the electric induction, magnetic field, electric field and magnetic 
induction four-vectors are covariantly defined as 
\begin{equation} 
D^i = H^{ik} U_k\,, \quad 
H^i = H^{*ik} U_k\,, \quad 
E^i = F^{ik} U_k\,, \quad 
B^i = F^{*ik} U_k\,, 
\label{constitutivevectors} 
\end{equation} 
respectively. 
These vectors are spacelike and orthogonal to the velocity 
four-vector $U^i$, 
 
\begin{equation} 
D^i U_i = 0 = E^i U_i, \quad H^i U_i = 0 = B^i U_i \, . 
\label{orthogonality1} 
\end{equation} 
For $U^i$ timelike and normalized to unity, one can represent the Maxwell 
tensor in the form 
\begin{equation} 
F_{mn} = \delta^{pq}_{mn} \ E_p U_q  - \epsilon_{mnls} \sqrt{-g} \  B^l U^s = 
E_m U_n - E_n U_m - \eta_{mnl} B^l \,, 
\label{fdecomposition} 
\end{equation} 
where $\delta^{ik}_{mn}$ is the generalized 4-indices Kronecker delta, 
and $\eta_{mnl}$ is an anti-symmetric tensor orthogonal to $U^i$ 
defined as 
\begin{equation} 
\eta_{mnl} \equiv \sqrt{-g} \ \epsilon_{mnls} \  U^s \,, 
\quad 
\eta^{ikl} \equiv \frac{1}{\sqrt{-g}} \ \epsilon^{ikls} \  U_s \,. 
\label{3levicivita} 
\end{equation} 
The tensor $\eta_{mnl}$ is a covariant generalization of three 
dimensional Levi-Civita symbol $e^{abc}$.  The identity 
\begin{equation} 
\frac{1}{2} \eta^{ikl}  \eta_{klm} = - \delta^{ij}_{ms} U_j U^s \, 
\label{identity2} 
\end{equation} 
will be important in what follows. 
 
Analogously to the $F_{mn}$ decomposition (\ref{fdecomposition}) one can 
represent the induction tensor $H_{ik}$ in the form 
\begin{equation} 
H_{ik} = \delta^{pq}_{ik} \ D_p U_q  - \epsilon_{ikls} \sqrt{-g} \ H^l U^s = 
D_i U_k - D_k U_i - \eta_{ikl} H^l \,. 
\label{ddecomposition} 
\end{equation} 
Taking into account the following covariant form of the relations 
(\ref{3model2}) 
\begin{equation} 
D^i = E^i + a \ \eta^{iml} \partial_m E_l \,, \quad 
H^i = B^i + b \ \eta^{iml} \partial_m B_l \,, 
\label{h3} 
\end{equation} 
as well as using the definitions (\ref{constitutivevectors}), 
we obtain from (\ref{fdecomposition}) and (\ref{ddecomposition}) the 
covariant constitutive equations for the two-parameter model 
\begin{equation} 
H^{ik} = F^{ik} + \frac{1}{2} \delta_{pq}^{ik} \  \nabla_s 
\left\{ \frac{\epsilon^{qjmn}}{\sqrt{-g}} \ U_j F_{mn} 
\left[a U^p U^s + b \Delta^{ps} \right] \right\} \,. 
\label{41fedorov} 
\end{equation} 
Comparing (\ref{constitutive}) and (\ref{41fedorov}) one sees that 
the $D^{iksmn}$ tensor takes the explicit form 
\begin{equation} 
D^{iksmn} = 
\frac{1}{2} \delta_{pq}^{ik} \left\{ \frac{\epsilon^{qjmn}}{\sqrt{-g}} \ U_j 
\left[a U^p U^s + b \Delta^{ps} \right] \right\} \,. 
\label{dtensor} 
\end{equation} 
Here $\Delta^{ps} \equiv g^{ps} - U^p U^s$ is a projector. 
 
\section{Curvature generalization of the two-parameter model of optical 
activity} 
 
\subsection{Generalization of the model} 
 
The curvature generalization of the model of optical activity can be 
obtained by replacing in the relation (\ref{41fedorov}) the 
Levi-Civita tensor $ \epsilon^{qjmn}/\sqrt{-g}$ by the dual Riemann 
tensor 
\begin{equation} 
R^{* qjmn} \equiv 
\frac{1}{2} R^{qj}_{\cdot \cdot \ ls} \ \frac{\epsilon^{lsmn}}{\sqrt{-g}}\,. 
\label{dualriemann} 
\end{equation} 
This replacement can be motivated as follows. Hehl and Obukhov \cite{hehl} 
state that a good linear ansatz for the constitutive relation between 
the induction tensor $H_{ij}$ and the Maxwell tensor $F_{kl}$ is 
\begin{equation} 
H_{ij}=F_{ij}+ \sqrt{-g} \ \epsilon_{ijmn}\chi^{mnkl}F_{kl} \,, 
\label{constitutive1} 
\end{equation} 
where $\chi^{mnkl}$ is a constitutive tensor with the same symmetries 
as the Riemann tensor. Now, the work of Drummond and Hathrell 
\cite{drummond}, on the corrections to quantum electrodynamics induced 
by curvature, tells us that a natural candidate to $\chi^{mnkl}$ is 
precisely the Riemann tensor $R^{mnkl}$. Thus we can replace one for 
the other in (\ref{constitutive1}) and obtain 
\begin{equation} 
H_{ij}=F_{ij}+ \chi R^{* \ kl}_{ij \cdot \cdot } F_{kl} \,, 
\label{constitutive2} 
\end{equation} 
where $\chi$ is some phenomenological coefficient. 
Let us summarize the following ideas discussed above: 
(i) the idea that the one-loop gravitationally corrected Maxwell 
equations are mathematically equivalent to the classical electrodynamic 
equations in the presence of a linear, anisotropic, non-homogeneous 
media, is a generally accepted idea; 
(ii)  the phenomenological approach to the modifications of 
Maxwell equations plays an important auxiliary role in an 
analysis of the effects induced by curvature; 
(iii) when the spacetime curvature influences the photons as a 
sort of anisotropic, non-homogeneous, non-stationary medium, one 
can expect that there exist the analogues of the corresponding 
phenomena in classical electrodynamics, indeed, the optical activity, 
as a well-known effect of linear dispersion, has to be considered among them; 
(iv) formally, the models describing the effects induced by 
curvature can be formulated by the following way: the tensor 
coefficients from the classical constitutive equations have to be 
replaced by the tensor coefficients containing the Riemann tensor.

Following this line, we can take the covariant constitutive equations 
(\ref{41fedorov}), describing the classical optical activity, and 
replace $\epsilon^{qjmn}/\sqrt{-g}$ by the dual Riemann tensor 
$R^{*qjmn}$. This procedure yields 
\begin{equation} 
H^{ik} = F^{ik} + \frac{1}{2} \delta_{pq}^{ik} \  \nabla_s 
\left\{ R^{*qjmn} \ U_j F_{mn} 
\left[ \bar{a} U^p U^s + \bar{b} \Delta^{ps}  \right] \right\} \,, 
\label{rfedorov} 
\end{equation} 
where now the gyration parameters $a$ and $b$ have been replaced 
by new phenomenological curvature gyration 
parameters $\bar a$ and $\bar b$, respectively. 
The units of $\bar a$ and $\bar b$ are $a \cdot{\rm length}^2$ and 
$b \cdot {\rm length}^2$. To simplify the notation we drop 
the bars from now on on $a$ and $b$. 
Now the Maxwell equation (\ref{maxwell1}) takes the form 
\begin{equation} 
\nabla_k F^{ik} + \frac{1}{2} \delta_{pq}^{ik} \ \nabla_k \nabla_s 
\left\{ R^{*qjmn} \ U_j F_{mn} 
\left[a U^p U^s + b \Delta^{ps}  \right] \right\} = 0 \,. 
\label{modifeq} 
\end{equation} 
As usually, we represent the Maxwell tensor $F_{ik}$ by 
\begin{equation} 
F_{ik} = \nabla_i A_k - \nabla_k A_i \,, 
\label{potential} 
\end{equation} 
where $A_i$ is the four-vector potential. With this definition 
equation (\ref{maxwell2}) is trivially satisfied. 
 
\subsection{Equations for optical activity induced by a gravitational pp-wave} 
 
The previous sections contain the description of the model for an 
arbitrary spacetime metric.  It is of interest to study this model in 
a particular background. The pp-wave solution 
is a  background well suited for our purposes, 
and incidentally has attracted much attention in the literature, 
notably in the gravitational fields generated within string theory, 
due to the fact that it is the Penrose limit of any spacetime 
\cite{penrose}.  The pp-wave,  a gravitational wave,  is an exact solution 
of General Relativity and has an explicit time dependence. 
The pp-wave is described by the metric \cite{mtw} 
\begin{equation} 
ds^{2} =  2 du dv - 
L^{2} \left[e^{2\beta}(dx^{2})^{2} + e^{-2\beta}(dx^{3})^{2} \right], 
\label{metric} 
\end{equation} 
where 
\begin{equation} 
u =  \frac{ct-x^{1}}{\sqrt{2}}\,, \quad  v = \frac{ct+x^{1}}{\sqrt{2}} \,, 
\label{time} 
\end{equation} 
are the retarded and the advanced time, respectively. The functions 
$L$ and $\beta$ depend on $u$ only, and are coupled 
by the equation 
\begin{equation} 
L^{''} + L \ (\beta^{'})^{2} = 0\,. 
\label{einstein} 
\end{equation} 
A prime denotes generically the derivative of the function with 
respect to its own argument, in particular here the derivatives of 
$L(u)$ and $\beta(u)$  with respect to the retarded time $u$. 
One can assume $\beta(u)$  as an arbitrary function of $u$ and 
then solve for $L$. 
The curvature tensor has  two  non-zero  components 
\begin{equation} 
- R^{2}_{\cdot u2u} = R^{3}_{\cdot u3u} = 
L^{-2} \left[L^{2}  \beta^{'} \right]^{'} \,. 
\label{riemann} 
\end{equation} 
As for the dual Riemann tensor, it has also two nonvanishing components 
\begin{equation} 
R^*_{2u3u} = R^*_{3u2u}= L^2 \  R^{3}_{\cdot u3u} \,. 
\label{starriemann} 
\end{equation} 
The Ricci tensor $R_{ik}$ and the curvature scalar $R$ are equal to zero. 
 
For a medium (or an observer)  at rest in the chosen frame of 
reference one has 
\begin{equation} 
U^i = (\delta^i_u + \delta^i_v) \frac{1}{\sqrt{2}}\, . 
\label{velocity} 
\end{equation} 
Calculating the components of the induction tensor $H_{ik}$ 
(\ref{rfedorov}) one obtains 
\begin{equation} 
H_{uv} = F_{uv} \,, 
\label{h1} 
\end{equation} 
\begin{eqnarray} 
& 
H_{v\sigma}= F_{v\sigma} + \frac{a+b}{2\sqrt2} \ \partial_v \left( 
R^*_{\sigma u \rho u} g^{\rho \gamma}  F_{\gamma v} \right) + 
\nonumber \\& 
+  \frac{a-b}{2\sqrt2} \left[\partial_u \left( 
R^*_{\sigma u \rho u} g^{\rho \gamma} F_{\gamma v} \right) 
+ \Gamma^\lambda_{\lambda u} R^*_{\sigma u \rho u} g^{\rho \gamma} 
F_{\gamma v} - 
\Gamma^\gamma_{\sigma u} R^*_{\gamma u \rho u} g^{\rho \lambda} 
F_{\lambda v}  \right] \,, & 
\label{hv2} 
\end{eqnarray} 
\begin{eqnarray} 
& 
H_{u\sigma}= F_{u\sigma} + \frac{a+b}{2\sqrt2} \ \partial_u \left( 
R^*_{\sigma u \rho u} g^{\rho \gamma}  F_{\gamma v} \right) + 
\nonumber \\& 
+  \frac{a-b}{2\sqrt2} \left[\partial_v \left( 
R^*_{\sigma u \rho u} g^{\rho \gamma} F_{\gamma v} \right) 
+ \Gamma^\lambda_{\lambda u} R^*_{\sigma u \rho u} g^{\rho \gamma} 
F_{\gamma v} - 
\Gamma^\gamma_{\sigma u} R^*_{\gamma u \rho u} g^{\rho \lambda} 
F_{\lambda v}  \right] \,, & 
\label{hu3} 
\end{eqnarray} 
\begin{equation} 
H_{23}= F_{23} + \frac{b}{2\sqrt2} g^{\gamma\rho} 
\left[ R^*_{3 u \gamma u} \partial_2 F_{\rho v} 
- R^*_{2 u \gamma u} \partial_3 F_{\rho v} \right] \,, 
\label{h4} 
\end{equation} 
where Greek indices take the values $2$ and $3$ only, 
and $\Gamma^i_{kl}$ are the Christoffel symbols. 
These $H_{ik}$ satisfy the following equations (see (\ref{maxwell1}) for 
the metric (\ref{metric})) 
\begin{equation} 
\partial_v H_{vu} + g^{\gamma \sigma} \partial_\gamma H_{v \sigma}=0 \,, 
\label{eq1} 
\end{equation} 
\begin{equation} 
\frac{1}{L^2} \partial_u \left( L^2 H_{uv} \right) + 
g^{\gamma \sigma} \partial_\gamma H_{u \sigma} =0 \,, 
\label{eq2} 
\end{equation} 
\begin{equation} 
\frac{1}{L^2}\partial_u \left( L^2 g^{\alpha \beta} H_{\beta v} \right) + 
g^{\alpha \beta} \partial_v H_{\beta u} + 
g^{\alpha \beta} g^{\gamma \sigma} \partial_\gamma H_{\beta \sigma} =0 \,. 
\label{eq34} 
\end{equation} 
Let us focus now on specific exact solutions of the Maxwell equations given 
by (\ref{h1})-(\ref{eq34}). 
 
\subsection{Solutions to the modified Maxwell equations} 
 
Consider electromagnetic waves 
traveling along the $x^1$ axis. 
For such a case the general solution of the electromagnetic potential $A_i$ 
can be written as 
\begin{equation} 
A_2= A_2(u,v)\,, \quad A_3 = A_3(u,v)\,, \quad A_u =0 \,, \quad A_v =0 \,. 
\label{a2a3} 
\end{equation} 
This means that the non-zero components of the Maxwell tensor are, 
\begin{equation} 
F_{u \alpha} = \partial_u A_\alpha \,, \quad 
F_{v \alpha} = \partial_v A_\alpha \,. 
\label{f23} 
\end{equation} 
For this case the equations (\ref{eq1}), (\ref{eq2}) become trivial. 
Introducing the new functions $M_2(u,v)$ and $M_3(u,v)$ by 
\begin{equation} 
A_2 = e^{\beta} M_2(u,v)\,, \quad 
A_2 = e^{-\beta} M_3(u,v) \,, 
\label{atob} 
\end{equation} 
and after some manipulations, one can reduce the two components in 
(\ref{eq34}) to the two following coupled equations 
\begin{eqnarray} 
& 
\partial_u \partial_v \left[ M_2 + \frac{a+b}{2\sqrt2} \ 
\partial_v \left( R^3_{\cdot u3u} M_3 \right) \right] + 
\nonumber \\& 
 \frac{a{-}b}{4\sqrt2} \left[ 
\left(\partial^2_u {+} \partial^2_v \right) {+} 
\left(\frac{L'}{L} {-} \beta' \right) \left(\partial_u {+} \partial_v \right) 
{-} \left((\beta')^2 {+} \left(\frac{L'}{L}\right)^2 {+} 
\beta'\frac{L'}{L} \right) 
\right] \partial_v  \left( R^3_{\cdot u3u} M_3 \right) {=} 0 \,, 
& 
\label{dalamber2} 
\end{eqnarray} 
\begin{eqnarray} 
& 
\partial_u \partial_v \left[ M_3 + \frac{a+b}{2\sqrt2} \ 
\partial_v \left( R^3_{\cdot u3u} M_2 \right) \right] + 
\nonumber \\& 
\frac{a{-}b}{4\sqrt2} \left[ 
\left(\partial^2_u {+} \partial^2_v \right) {+} 
\left(\frac{L'}{L} {+} \beta' \right) \left(\partial_u {+} \partial_v \right) 
{-} \left((\beta')^2 {+} \left(\frac{L'}{L}\right)^2 {-} \beta'\frac{L'}{L} 
\right) \right] \partial_v  \left( R^3_{\cdot u3u} M_2 \right) 
{=} 0 \,. 
& 
\label{dalamber3} 
\end{eqnarray} 
Note that the equation (\ref{dalamber3}) can be obtained from (\ref{dalamber2}) 
by interchanging the indices 2 and 3, as well as $\beta$ with $-\beta$. 
 
Now, a simple model is obtained when $a=b$. In this case, the second 
part in equations (\ref{dalamber2}) and (\ref{dalamber3}) are zero, 
one has a D'Alembertian operator alone, and the model is integrable in 
terms of elementary functions.  The other cases are represented 
generically by $a\neq b$. We will discuss two interesting particular 
cases, the Landau-Lifshitz model with $b=0$, and the Fedorov model 
with $a=-b$. 
 
\subsubsection{Exact integrable one-parameter model with $a=b$} 
 
In the framework of this one-parameter model equations (\ref{dalamber2}) and 
(\ref{dalamber3}) reduce to two D'Alembert type equations 
\begin{equation} 
\partial_u \partial_v \left[ M_2 + Q(u) \ \partial_v M_3 \right] =0 \,, 
\label{dalamber21} 
\end{equation} 
\begin{equation} 
\partial_u \partial_v \left[ M_3 + Q(u) \ \partial_v M_2 \right] =0 \,, 
\label{dalamber31} 
\end{equation} 
where 
\begin{equation} 
Q(u) \equiv \frac{a}{\sqrt{2} L^2} R^*_{2u3u} =  \frac{a}{\sqrt2} 
R^3_{\cdot u3u} \,. 
\label{e} 
\end{equation} 
Solving equations (\ref{dalamber21}) and (\ref{dalamber31}) we obtain 
\begin{equation} 
M_2(u,v) + Q(u) \ \partial_v M_3(u,v) = \Phi_2(u) + \Psi_2(v) \,, 
\label{int2} 
\end{equation} 
\begin{equation} 
M_3(u,v) + Q(u) \ \partial_v M_2(u,v)  = \Phi_3(u) + \Psi_3(v) \,, 
\label{int3} 
\end{equation} 
where $\Phi_\alpha(u)$ and $\Psi_\alpha(v)$ are arbitrary functions of 
their arguments.  The standard procedure of the integration of such a 
system is the following.  First, one substitutes the $M_2(u,v)$ 
function given in equation (\ref{int2}) into equation (\ref{int3}), 
and obtains a second order differential equation for $M_3(u,v)$, 
\begin{equation} 
M_3(u,v) - Q^2(u) \ \partial^2_v M_3(u,v)  = \Phi_3(u) + \Psi_3(v) - 
Q(u) \ \partial_v \Psi_2(v) \,. 
\label{twodim} 
\end{equation} 
Since the coefficients in (\ref{twodim}) depend on $u$ only, one can 
use the Fourier transforms of $M_\alpha(u,v)$ and other functions 
defined below, in order to find the solutions of the respective equation. 
The relation between a Fourier transform quantity ${\cal T}(u,k)$ and 
the function $T(u,v)$ is given by 
\begin{equation} 
T(u,v) = \int_{-\infty}^{\infty} \frac{d k}{2\pi} {\cal T}(u,k) e^{-ikv} 
\,. 
\label{four} 
\end{equation} 
Applying a Fourier transformation to the quantity $M_3(u,v)$, one obtains 
from equation (\ref{twodim}), for the Fourier transform ${\cal M}_3(u,k)$, 
the formula 
\begin{equation} 
{\cal M}_3(u,k) = \frac{{\cal F}(u,k)}{1+k^2 Q^2(u)} 
\,, 
\label{bfour} 
\end{equation} 
where 
\begin{equation} 
{\cal F}(u,k) = \int_{-\infty}^{\infty} d v e^{ikv} \left[ 
\Phi_3(u) + \Psi_3(v) - Q(u) \ \partial_v \Psi_2(v) \right] \,. 
\label{psifour} 
\end{equation} 
This is valid for arbitrary initial functions 
$\Phi_\alpha(u)$ and $\Psi_\alpha(v)$. Let us consider now a 
special set of initial functions, which yield an exact solution 
in terms of elementary functions. We choose $\Psi_\alpha(v)$ as 
\begin{equation} 
\Psi_2(v) = \Psi_2(0) \cos{(kv + \varphi_0)} \,, \quad 
\Psi_3(v) = \Psi_3(0) \sin{(kv + \varphi_0)} \,. 
\label{cosin} 
\end{equation} 
We also search for $M_\alpha(u,v)$ in the form 
\begin{equation} 
M_\alpha(u,v)  = \Phi_\alpha(u) + X_\alpha(u) \ \Psi_\alpha(v)\,. 
\label{search} 
\end{equation} 
The functions $X_\alpha(u)$ have to be found from the equations 
(\ref{int2}) and (\ref{int3}). These equations yield 
\begin{equation} 
\frac{1 - X_2(u)}{Q(u) X_3(u)} = \frac{\Psi_3'(v)}{\Psi_2(v)} = 
k \frac{\Psi_3(0)}{\Psi_2(0)} \,, 
\label{const1} 
\end{equation} 
and 
\begin{equation} 
\frac{1 - X_3(u)}{Q(u) X_2(u)} = \frac{\Psi_2'(v)}{\Psi_3(v)} = 
- k \frac{\Psi_2(0)}{\Psi_3(0)} \,, 
\label{const2} 
\end{equation} 
where the right hand side in equations (\ref{const1}) and (\ref{const2}) 
were put equal to a constant of separation of variables, as usual in 
partial differential equations. These constants were chosen in 
accord the initial conditions given in equation (\ref{cosin}). 
The solutions to  $X_\alpha(u)$ can be obtained from (\ref{const1}) and 
(\ref{const2}) and have the form 
\begin{equation} 
X_2(u) = \frac{\Psi_2(0) - k Q(u) \Psi_3(0)}{\Psi_2(0) (1 + k^2 Q^2(u))}\,, 
\quad 
X_3(u) = \frac{\Psi_3(0) + k Q(u) \Psi_2(0)}{\Psi_3(0) (1 + k^2 Q^2(u))}\,. 
\label{x2x3} 
\end{equation} 
Thus, from (\ref{atob}), (\ref{search}), and (\ref{x2x3}), we obtain for the 
potentials 
\begin{equation} 
A_2 = e^{\beta} \left[ \Phi_2(u) + 
\frac{\Psi_2(0) - k Q(u) \Psi_3(0)}{1 + k^2 Q^2(u)} \cos{(kv + \varphi_0)} 
\right] \,, 
\label{sola2} 
\end{equation} 
\begin{equation} 
A_3 = e^{- \beta} \left[ \Phi_3(u) + 
\frac{\Psi_3(0) + k Q(u) \Psi_2(0)}{1 + k^2 Q^2(u)} \sin{(kv + \varphi_0)} 
\right] \,. 
\label{sola3} 
\end{equation} 
The electromagnetic field represented by the formulae (\ref{sola2}) and 
(\ref{sola3}) consists of a mixture of electromagnetic waves traveling 
in the same direction and in opposite direction to the gravitational 
pp-wave direction. 
 
For $\Psi_\alpha(v)=0$, the electromagnetic wave is a function of the 
retarded time $u$ only, and therefore travels in the direction of the 
pp-wave. In this case there is no optical activity (see, (\ref{sola2}) and 
(\ref{sola3})), since the electromagnetic wave does not couple with the 
curvature (see, also, \cite{bala,bale1,bale2}). 
 
Thus, let us study the case of electromagnetic waves traveling in an 
opposite 
direction to the gravitational wave. First, we assume 
\begin{equation} 
\Phi_2(u) = \Phi_3(u)= 0 \,. 
\label{simple} 
\end{equation} 
Then, eliminating the variable $v$ in equations (\ref{sola2}) and 
(\ref{sola3}) one obtains the equation for the 
ellipse of polarization 
\begin{equation} 
\frac{( A_2)^2 }{( e^{\beta}X_2(u))^2} + 
\frac{(A_3)^2}{( e^{-\beta}X_3(u))^2} = 1 \,, 
\label{ellips} 
\end{equation} 
where the quantities $e^{\beta}X_2(u)$ and $e^{-\beta}X_3(u)$ play the role 
of semi-axes for 
this ellipse. The instantaneous position of the polarization vector can be 
characterized by the angle $\varphi(u,k)$, which satisfy the relationship 
\begin{equation} 
\tan{\varphi(u,k)} = \frac{X_3(u)}{X_2(u)} = 
\frac{\Psi_3(0) + k Q(u) \Psi_2(0)}{\Psi_2(0) - k Q(u) \Psi_3(0)} \,. 
\label{rotangle} 
\end{equation} 
We see, that the polarization vector is rotating. The frequency of rotation, 
$\nu(u,k)$, is equal to 
\begin{equation} 
\nu(u,k) \equiv \frac{d}{ds}\varphi(u,k)= 
\frac{\partial}{\partial u}\varphi(u,k) \ \frac{du}{ds} 
=\frac{\partial}{\partial u}\varphi(u,k) \ U^u = 
\frac{1}{\sqrt2}\frac{\partial}{\partial u}\varphi(u,k) \,. 
\label{nudef} 
\end{equation} 
Thus, 
\begin{equation} 
\nu(u,k) = \frac{ k Q^{'}(u)}{\sqrt2(1 + k^2 Q^2(u))} \,. 
\label{rotfreq} 
\end{equation} 
The initial values $\Psi_2(0)$ and $\Psi_3(0)$ are arbitrary. 
Defining $\omega_{\rm em}$ as the initial frequency of the 
electromagnetic wave, one has $k= \omega_{\rm em}/c$.  From 
(\ref{e}), $Q(u)$ is proportional to the Riemann tensor component. 
Thus, from equation (\ref{rotfreq}), we find that there is rotation of the 
polarization vector. This means there is a net Faraday effect,  
induced by curvature in this instance.

\subsubsection{Solutions for  $a \neq b$} 
 
For $a\neq b$ there are many possible models. We select two. 
\bigskip 
 
\noindent {\bf (i)} {\bf The Landau-Lifshitz model: $b=0$} 
 
\noindent In contrast to the case $a=b$, the models with $a\neq b$ do 
not admit a representation of the solution in terms of elementary 
functions.  However, for a small enough ratio of $\omega_{\rm 
GW}/\omega_{\rm em}$, where $\omega_{\rm GW}$ is the 
gravitational  wave frequency, 
one can apply the Fourier transformation (\ref{four}) to 
the equations (\ref{dalamber2}) and (\ref{dalamber3}).  This is a case 
that occurs in astrophysical situations of interest (see, e.g, 
\cite{mtw,thorne})). 
Putting  $b=0$ in equations (\ref{dalamber2}) and (\ref{dalamber3}) 
and taking into account the leading terms in 
$k=\omega_{\rm em}/c$, 
one obtains the following equations for the Fourier transforms 
${\cal M}_2(u,k)$ and ${\cal M}_3(u,k)$ 
\begin{equation} 
\partial_u {\cal M}_2(u,k) =  \frac{k^2 a}{4\sqrt2} R^3_{\cdot u3u} \ 
{\cal M}_3(u,k) \,, 
\quad 
\partial_u {\cal M}_2(u,k) =  \frac{k^2 a}{4\sqrt2 } R^3_{\cdot u3u} \ 
{\cal M}_2(u,k) \,. 
\label{foureq} 
\end{equation} 
The solution of (\ref{foureq}) is 
\begin{eqnarray} 
& 
{\cal M}_2(u,k) = 
{\cal M}_2(0,k) \cosh{\Theta(u,k)} + {\cal M}_3(0,k) \sinh{\Theta(u,k)} \,, 
\nonumber \\& 
{\cal M}_3(u,k) = 
{\cal M}_3(0,k) \cosh{\Theta(u,k)} + {\cal M}_2(0,k) \sinh{\Theta(u,k)} \,, 
& 
\label{foursol} 
\end{eqnarray} 
where 
\begin{equation} 
\Theta(u,k) \equiv  \frac{ k^2 a}{4\sqrt2} \int_0^u 
du R^3_{\cdot u3u}(u) \,. 
\label{theta} 
\end{equation} 
One sees from (\ref{foursol}) and (\ref{theta}) that the plane of 
polarization of each Fourier component, with fixed $k \neq 0$, is 
rotating, the rotation being hyperbolic. A particular case, worth of 
comment, is when the electromagnetic wave is polarized at $u=0$ in the 
$x^2-$direction, and the Fourier amplitude ${\cal M}_3(0,k)$ is set to 
zero, ${\cal M}_3(0,k) =0$, also at $u=0$.  Then, from equation 
(\ref{foursol}), one finds that the ${\cal M}_3(u,k)$ component 
becomes nonvanishing at $u>0$, since we have assumed 
${\cal M}_2(0,k)\neq 0$.  In this particular case, the angle of rotation 
$\varphi(u,k)$, defined by 
$\tan{\varphi(u,k)} \equiv{\cal M}_3(u,k)/{\cal M}_2(u,k)$, is equal to 
$\varphi(u,k) = \arctan{(\tanh{\Theta(u,k)})}$. The corresponding frequency 
of the rotation of the polarization plane, $\nu(u,k)$ has the form 
\begin{equation} 
\nu(u,k) = 
\frac{\frac{\partial}{\partial u}\Theta(u,k)}{\sqrt2\cosh{2\Theta(u,k)}} \,. 
\label{llrot} 
\end{equation} 
 
\bigskip 
\noindent {\bf (ii)} {\bf The Fedorov model: $a=-b$} 
 
\noindent In classical electrodynamics the Fedorov model 
possesses the same duality 
invariance, as the Maxwell equations (\ref{3max}). Applying the 
transformation (see, e.g., \cite{jackson}) 
\begin{eqnarray} 
& 
\vec{E} = \vec{E}' \cos{\xi} + \vec{H}' \sin{\xi} \,, \quad 
\vec{H} = - \vec{E}' \sin{\xi} + \vec{H}' \cos{\xi}  \,, 
\nonumber \\& 
\vec{D} = \vec{D}' \cos{\xi} + \vec{B}' \sin{\xi} \,, \quad 
\vec{B} = - \vec{D}' \sin{\xi} + \vec{B}' \cos{\xi} 
& 
\label{jackson} 
\end{eqnarray} 
with arbitrary constant $\xi$ to the Maxwell equations (\ref{3max}), 
one obtains the same equations for the quantities with 
primes. Analogously, applying the transformations (\ref{jackson}) to 
the equations (\ref{3model}) one sees that these equations are duality 
invariant when $a=-b$ only, i.e. in the Fedorov model. 
 
Let us consider now the curvature generalization of the Fedorov model 
and solve the equations (\ref{dalamber2}) and (\ref{dalamber3}) in the 
case $a = -b$. In a leading order approximation in $k=\omega_{\rm 
em}/c$ (see part (i) of this subsubsection) one can see that the 
solutions coincide with the solutions describing the Landau-Lifshitz 
model, but instead of $a$ one has to insert $2a$.  In other words, 
this model also describes a Faraday rotation of the polarization plane 
with double rotation frequency as comparison to the model $b=0$ (see 
equation (\ref{llrot})).

\subsubsection{Orders of magnitude for the Faraday rotation } 
 
It is interesting to compare the frequencies of rotation of the 
polarization vector in the model $a=b$ on one hand, and the models 
$b=0$, $a=-b$ on the other hand, when $a R^3_{\cdot u3u}$ is small. 
In this case, formula (\ref{rotfreq}) for the model $a=b$ 
yields a frequency 
\begin{equation} 
\nu_{\rm a=b} = \frac{1}{2} ka (R^3_{\cdot u3u}(u))' \,. 
\label{f1} 
\end{equation} 
For the Landau and Lifshitz model formula (\ref{llrot}) 
gives 
\begin{equation} 
\nu_{\rm b=0} =\frac{1}{8} k^2 a R^3_{\cdot u3u}(u) \,, 
\label{f2} 
\end{equation} 
and for 
the Fedorov model one finds 
\begin{equation} 
\nu_{\rm a=-b} = 2 \ \nu_{\rm b=0}\,. 
\label{f3} 
\end{equation} 
One sees that the ratios $\nu_{\rm b=0} / \nu_{\rm a=b}$ and 
$\nu_{\rm a=-b} / \nu_{\rm a=b}$ are of the same order of magnitude as 
the ratio of $\omega_{{\rm em}} /\omega_{{\rm GW}}$.  In other words, for 
$\omega_{{\rm em}} /\omega_{{\rm GW}} >> 1$, the rate of rotation of 
the polarization vector induced by curvature, predicted in the 
Landau-Lifshitz and Fedorov models, is much faster than the rotation 
rate predicted in the model with $a=b$. 
 
Let us estimate numerically the typical magnitude of the effect of the 
rotation of the polarization vector induced by the curvature 
on the example of 
the Fedorov model. One obtains from (\ref{theta}), (\ref{riemann}) 
and (\ref{f3}) 
that in the weak gravity field approximation 
\begin{equation} 
\Theta(u^*,k) =  \frac{ k^2 a}{2\sqrt2} \left[\beta^{\prime}(u^*) - 
\beta^{\prime}(0) \right] \,. 
\label{approx} 
\end{equation} 
Consider now the simple model in which the electromagnetic signal 
emitted by some astrophysical source passes 
through the environment of the pulsar PSR J0534+2200 in the CRAB 
nebula and arrives at the Earth. 
The parameters of this pulsar are well-known (e.g. \cite{pizz}): 
the period of rotation is $T_{{\rm rot}} = 33.4\, {\rm ms}$, the 
frequency of the 
gravitational radiation is $\nu_{{\rm GW}}=60\, {\rm Hz}$, the distance is 
equal to $d = 6.2\times 10^{21} \,{\rm cm}$, 
and the optimistic estimation 
of the amplitude of the gravitational wave on the Earth surface is 
$h = 9.5 \times 10^{-25}$. 
 
One can consider the 
moment of the retarded time $u=u^*$ as the moment of the arrival of the 
signal on Earth. As for 
$u=0$, it can be considered as the moment when the electromagnetic 
signal passes in the vicinity of the pulsar. 
Taking into account that the amplitude 
of the gravitational wave decreases reciprocally to the distance from 
the source, and the gravitational wave field is modeled to be periodic 
with the frequency $\omega_{\rm GW}= 2 \pi \nu_{{\rm GW}}$, 
one can represent the leading 
term in (\ref{approx}) in the following form 
\begin{equation} 
\vert \Theta(u^*) \vert = 
\frac{ a \ \beta_0 \ \omega^2_{{\rm em}} \ \pi \nu_{{\rm GW}} }{c^3} \,, 
\label{1approx} 
\end{equation} 
where $\beta_0$ is the amplitude of the gravitational wave at the 
minimal distance from the pulsar (it is of the order of 
the neutron star radius 
$R_0=10^{6}\,{\rm cm}$), and $ k\, c \,= \,\omega_{{\rm em}}$. 
Putting 
\begin{equation} 
\beta_0 = h \,\frac{d}{R_0} \simeq 5.9 \times 10^{-9} \,, 
\label{33approx} 
\end{equation} 
we obtain the following estimate for the angle of rotation 
of the polarization vector induced by curvature 
\begin{equation} 
\vert \Theta(u^*) \vert \simeq 41 \, {\rm rad} \, 
\left(\frac{a}{1 \ {\rm m}^3} \right) 
\left(\frac{\beta_0}{5.9 \times 10^{-9}} \right) 
\left(\frac{\omega_{{\rm em}}/ 2\pi}{5 \times 10^{15}\,{\rm Hz} } 
\right)^2 
\left(\frac{\nu_{{\rm GW}}}{60\,{\rm Hz}} \right) \,. 
\label{2approx} 
\end{equation} 
In order to estimate the parameter $a$ one can mention that for 
a crystal of quartz the standard gyration coefficient $\gamma^{123}$ 
(see (\ref{3constitutive})) for the frequency $\nu_{{\rm em}} = 
5 \times 10^{15} \,{\rm Hz}$ 
is equal to $\gamma^{123} = 1.2 \times 10^{-10} \,{\rm cm}$ (see, e.g. 
\cite{sirotin}). This means that the ratio 
$\xi_{{\rm quartz}} \equiv \frac{\gamma^{123}}{\lambda_{{\rm em}}}$, where 
$\lambda_{{\rm em}}$ is the wavelength of the electromagnetic wave, 
is of the order of $10^{-6}\,$. 
Analogously, we can introduce a new dimensionless parameter 
$\xi_{{\rm GW}}$ using, for example, the following relation 
\begin{equation} 
a = \xi_{{\rm GW}}\, \lambda_{{\rm em}} \,\lambda^2_{{\rm GW}} \simeq 
\xi_{{\rm GW}} \times 10^{13}\, {\rm cm}^3 \,, 
\label{2estima} 
\end{equation} 
where $\lambda_{{\rm GW}} \equiv \frac{c} {\nu_{{\rm GW}}} 
\simeq 5 \times 10^{8}\, {\rm cm} $ 
is the wavelength of gravitational radiation. 
Thus, the estimation for the angle rotation yields 
\begin{equation} 
\vert \Theta(u^*) \vert \simeq 41 \, {\rm rad} \times 
(10^7 \times \xi_{{\rm GW}})  \,, 
\label{4approx} 
\end{equation} 
where $\xi_{{\rm GW}} <<\xi_{{\rm quartz}} < 10^{-6}$ 
has to be estimated experimentally.

\section{Conclusions} 

We have established a model with two gyration parameters $a$ and $b$, 
for optical activity induced by curvature. Within this two-parameter 
model we have studied three models, namely, $a=b$, $b=0$, and $a=-b$. 
It was shown that the $a=b$ model is exactly integrable, while for the 
models $b=0$ (the Landau-Lifshitz model) and $a=-b$ (the Fedorov 
model) the solutions are presented in the short wavelength 
approximation.  The solutions show that the optical activity induced 
by curvature leads to Faraday rotation, i.e., the rotation of the 
plane of polarization of the electromagnetic wave traveling in the 
gravitational wave background.  The frequency of the rotation is 
deternd.  The frequency of the rotation is 
determined by the value of the Riemann tensor and by the 
phenomenological gyration parameters $a$ and $b$.

\vspace{1.5 cm} 
\noindent {\large\bf Acknowledgments} 
This work was partially funded by the Portuguese 
Science Foundation, FCT, through project PESO/PRO/2000/4014.

\end{document}